\documentclass[12pt]{article}

\begin{document}
\title{Doubly Special Relativity: facts and prospects}
\author{J.\ Kowalski--Glikman\thanks{e-mail
address jurekk@ift.uni.wroc.pl}\\ Institute for Theoretical
Physics\\ University of Wroclaw\\ Pl.\ Maxa Borna 9\\ Pl--50-204
Wroclaw, Poland} \maketitle

\begin{abstract}
In this short review of Doubly Special Relativity I describe first
the relations between DSR and (quantum) gravity. Then I show how, in
the case of a field theory with curved momentum space, the Hopf
algebra of symmetries naturally emerges. I conclude with some
remarks concerning DSR phenomenology and description of open
problems.
\end{abstract}

\section{Introduction: What is DSR?}

The definition of  Doubly Special Relativity (DSR)
(Amelino-Camelia, 2001 and 2002, see Kowalski-Glikman, 2005 for
review) is deceptively simple. Recall that Special Relativity is
based on two   postulates: Relativity Principle for inertial
observers and the existence of a single observer independent scale
associated with velocity of light. In this DSR replaces the second
postulate by assuming existence of {\em two} observer independent
scales: the old one of velocity plus the scale of mass (or of
momentum, or of energy). That's it.

 Adding new postulate has consequences, however.  The most immediate on is
 the question:  what does the second observer--independent scale mean physically?
  Before trying to answer this question, let us recall the concept of an
observer--independent scale. It  can be easily understood, when
contrasted with the notion of dimensionful coupling constant, like
Planck constant $\hbar$ or gravitational constant $G$. What is their
status in relativity? Do they transform under  Lorentz
transformation? Well, naively,  one would think that they should
because they are given by dimensional quantities. But of course they
do not. The point is that there is a special operational definition
of these quantities. Namely each observer, synchronized with all the
other observers, by means of the standard Einstein synchronization
procedure, measures their values in an identical quasi--static
experiment in her own reference frame (like Cavendish experiment).
Then the relativity principle assures that the numerical value of
such a constant will turn out to be the same in all experiments (the
observers could check the validity of relativity principle by
comparing values they obtained in their experiments). With an
observer independent scale the situation is drastically different.
Like the speed of light it cannot be measured in quasi-static
experiments; all the observers now measure a quantity associated
with a single object (in Special Relativity, all the observers could
find out what the speed of light is just by looking at the same
single photon.)

Now DSR postulates presence of the second observer--independent
scale. What is the physical  object that carries this scale, like
the photon carrying the scale of velocity of light? We do not know.
One can speculate that black hole remnants will do, but to
understand them we need, presumably, the complete theory of quantum
gravity. Fortunately, there is another way one can think of the
observer--independent scale. If such a scale is present in the
theory, and since, as explained above, it is operationally defined
in terms of experiments, in which one physical object is observed by
many distinct observers, who all measure the same value of the
scale, it follows that the scale must appear as a parameter in the
transformation rules, relating observers to each other. For example
velocity of light is present as a parameter in Lorentz
transformations. If we have a theory of spacetime with two observer
independent scales, both should appear in the transformations. As an
example one can contemplate the following form of infinitesimal
action of Lorentz generators, rotations $M_i$ and boosts $N_i$
satisfying the standard Lorentz algebra, on momenta (so called
DSR1), with the scale of mass $\kappa$
$$
[M_i, P_j] =  \epsilon_{ijk} P_k, \quad [M_i, P_0] =0
$$
$$
   \left[N_{i}, {P}_{j}\right] =   \delta_{ij}
 \left( {1\over 2} \left(
 1 -e^{-2{P_0}/\kappa}
\right) + {{\mathbf{P}^2}\over 2\kappa}  \right) -  \frac1\kappa\, P_{i}P_{j},\nonumber
$$
\begin{equation}\label{1}
  \left[N_{i},P_0\right] =  P_{i}.
\end{equation}
This algebra is a part of $\kappa$-Poincar\'e quantum algebra, see
Majid and Ruegg, 1994. One can also imagine a situation in which the
scale $\kappa$ appears not in the rotational, but in the
translational sector of the modified, deformed Poincar\'e group.

One may think of the second scale also in terms of synchronization
of observers. Recall   that the velocity of light scale is
indispensable in Special Relativity because it provides the only
meaningful way of synchronizing different observers. However this
holds for spacetime measurements (lengths and time intervals) only.
To define momenta and energy, one  must relate them to velocities.
On the other hand, using the momentum scale, one could, presumably,
make both the spacetime and momentum space synchronization,
independently, and perhaps could even describe the phase space as a
single entity. Thus it seems that in DSR the primary concept would
be the phase space not the configuration one.

In the limit when the second scale is very large (or very small
depending on how the  theory is constructed) the new theory should
reduce to the old one, for example when the second scale $\kappa$ of
DSR goes to infinity, DSR should reduce to Special Relativity.
Putting it another way we can think of DSR as some sort of
deformation of SR. Following this understanding some researchers
would translate the acronym DSR to Deformed Special Relativity. But
of course, deformation requires a deformation scale, so even
semantically both terms are just equivalent, just stressing
different aspects of DSR. Note that  in addition to the modified,
deformed algebra of spacetime symmetries, like the one in eq.\
(\ref{1}), the theory is to be equipped with an additional
structure(s), so as to make sure that its algebra cannot be reduced
to the standard algebra of spacetime symmetries of Special
Relativity, by rearrangement of generators. Only in such  a case DSR
will be physically different from Special Relativity.

In the framework of DSR we  want to understand if there are any
modifications  to the standard particle kinematics as described by
Special Relativity, at very high energies, of order of Planck scale.
The motivation is both phenomenological and theoretical. First there
are indications from observations of cosmic rays carrying energy
higher than GZK cutoff that the standard Special relativistic
kinematics might be not an appropriate description of particle
scatterings at energies of order of $10^{20}\, eV$ (in the
laboratory frame). Similar phenomenon, the violation of the
corresponding cutoff predicted by the standard Special Relativistic
kinematics for ultra-high energy photons seems also to be observed.
It should be noted however that in both these cases we do not really
control yet all the relevant astrophysical details of the processes
involved (for example in the case of cosmic rays we do not really
know what are the sources, though it it is  hard to believe that
they are not at the cosmological distances.) The extended discussion
of these issues can be found, for example, in Aloisio {\em et.\
al.}, 2005. If violation of the GZK cutoff is confirmed, and if
indeed the sources are at the cosmological distances, this will
presumably indicate the deviation from Lorentz kinematics. One of
the major goals of DSR is to work out the robust theoretical
predictions, concerning magnitude of such effects. I will briefly
discuss the ``DSR phenomenology'' below.

\section{Gravity as the origin of DSR}

The idea of DSR arose from the desire to describe possible
deviations from the  standard Lorentz kinematics on the one hand
and, contrary to the Lorentz breaking schemes, to preserve  the most
sacred principle of physics -- the relativity principle. Originally
the view was that one may be forced by phenomenological data to
replace Special Relativity by DSR, and then, on the basis of the
latter one should construct its  curved space extension, ``Doubly
General Relativity''. Then it has been realized that, in fact, the
situation is likely to be quite opposite: DSR might be {\em the}
correct flat space limit of gravity coupled to particles (see
Amelino-Camelia {\em et.\ al.}, 2004 and Freidel {\em et.\ al.},
2004)

We are thus facing the fundamental theoretical question: is Special
Relativity indeed, as it is believed, the correct limit of (quantum)
gravity in the case when spacetime is flat? From the perspective of
gravity flat Minkowski spacetime is some particular configuration of
gravitational field, and us such is to be described by  theory of
gravity. It corresponds to configurations of gravitational field in
which this field vanishes. However equations governing gravitational
field are differential equations and thus describe the solutions
only locally. In the case of Minkowski space particle kinematics we
have to do not only with (flat) gravitational field but also with
particles themselves. The particles are, of course, the sources of
gravitational field and even in flat space limit the trace of
particles' back reaction on spacetime might remain in the form of
some global information, even if locally, away from the locations of
the particle, the spacetime is flat. Of course we know that in
general relativity energy-momentum of matter curves spacetime, and
the strength of this effect is proportional to gravitational
coupling (Newton's constant.) Thus we are interested in the
situation, in which the transition from general relativity to
special relativity corresponds to smooth switching off the
couplings. In principle two situations are possible (in 4
dimensions):

\begin{enumerate}

\item weak gravity, semiclassical limit of quantum gravity:
\begin{equation}\label{2}
   G,\, \hbar \rightarrow 0, \quad \sqrt{\frac\hbar{G}} = \kappa \, \mbox{ remains finite}
\end{equation}

\item weak gravity, small cosmological constant limit of quantum gravity:
\begin{equation}\label{3}
   \Lambda \rightarrow 0, \quad  \kappa \, \mbox{ remains finite}
\end{equation}
\end{enumerate}

The idea is therefore to devise a controllable transition from the
full (quantum) gravity coupled to point particles to the regime, in
which all local degrees of freedom of gravity are switched off. Then
it is expected that locally, away from particles' worldlines gravity
will take the form of Minkowski (for $\Lambda=0$) or (Anti) De
Sitter space, depending on the sign of $\Lambda$. Thus it is
expected that DSR arises as a limit of general relativity coupled to
point particles in the topological field theory limit. To be more
explicit, consider the formulation of gravity as the constrained
topological field theory, proposed in Freidel and Starodubtsev,
2005.
\begin{equation}
S=\int \left(B_{IJ}\wedge F^{IJ}
-\frac{\alpha}{4}\, B_{IJ}\wedge B_{KL}\epsilon^{IJKL5}-  \frac{\beta}{2}\, B^{IJ}\wedge B_{IJ}\right) \label{4}.
\end{equation}
Here $F^{IJ}$ is the curvature of $SO(4,1)$ connection $A^{IJ}$, and $B_{IJ}$ is a two-form valued in the algebra $SO(4,1)$. The dimensionless parameters $\alpha$ and $\beta$ are related to gravitational and cosmological constants, and Immirzi parameter. The $\alpha$ term breaks the symmetry, and for $\alpha \neq 0$ this theory is equivalent to general relativity. On the other hand there are various limits in which this theory becomes a topological one. For example, for $\alpha \rightarrow 0$ all the local degrees of freedom of gravity disappear, and only the topological ones remain. One hopes that after coupling this theory to point particles, one derives DSR in an appropriate, hopefully natural, limit. This hope is based on experience with the 2+1 dimensional case, which I will now discuss.

\section{Gravity in 2+1 dimensions as DSR theory}

It is well known that gravity in 2+1 does not possess local degrees
of freedom and  is described by a topological field theory. Even in
the presence of point particles with mass and spin the 2+1
dimensional spacetime is locally flat. Thus 2+1 gravity is a perfect
testing ground for DSR idea. There is also a simple argument that it
is not just a toy model, but can tell us something about the full
3+1 dimensional case. It goes as follows.

As argued above what we are interested in is the flat space limit of
gravity  (perhaps also the semiclassical one in the quantum case.)
Now consider the situation when we have 3+1 gravity coupled to
planar configuration of particles. When the local degrees of freedom
of gravity are switched off this configuration has translational
symmetry along the direction perpendicular to the plane. But now we
can make the dimensional reduction and describe the system
equivalently with the help of 2+1 gravity coupled to the particles.
The symmetry algebra in 2+1 dimensions must be therefore a
subalgebra of the full 3+1 dimensional one. Thus if we find that the
former is not the 2+1 Poincar\'e algebra but some modification of
it, the latter must be some appropriate modification of the 3+1
dimensional Poincar\'e algebra. Thus if DSR is relevant in 2+1
dimensions, it is likely that it is going to be relevant in 3+1
dimensions as well.

Let us consider the analog of the situation 2, listed in the
previous section.  We start therefore with the 2+1 gravity with
positive cosmological constant. Then it is quite well well
established (see for example Noui and Roche, 2003) that the
excitations  of $3d$ quantum gravity with cosmological constant
transform under representations of the quantum deformed de Sitter
algebra $SO_q(3,1)$,  with $z=\ln q$ behaving in the limit of small
$\Lambda \hbar^2 /\kappa^2$ as $z  \approx  \sqrt{\Lambda} \hbar /
\kappa$, where $\kappa$ is equal to inverse 2+1 dimensional
gravitational constant, and has dimension of mass.

I will not discuss at this point the notion of quantum deformed
algebras  (Hopf algebras) in much details  It  suffices to say that
quantum algebras consist of several structures, the most important
for our current purposes would be the universal enveloping algebra,
which could be understand as an algebra of brackets among
generators, which are equal to some analytic functions of them. Thus
the quantum algebra is a generalization of a Lie algebra, and it is
worth observing that the former reduces to the latter in an
appropriate limit. The other structures of Hopf algebras, like
co-product and antipode, are also relevant in the context of DSR,
and I will introduce them in the next section.

 In the case of quantum algebra  $SO_q(3,1)$ the algebraic part looks  as follows  (the parameter
 $z$ used below is related to $q$ by $z=\ln q$)
\begin{eqnarray}
&& [M_{2,3},M_{1,3}]  =  {1 \over z} \sinh(z M_{1,2}) \cosh(z M_{0,3})
\nonumber \\
&& [M_{2,3},M_{1,2}] = M_{1,3}
\nonumber \\
&& [M_{2,3},M_{0,3}] = M_{0,2}
\nonumber \\
&& [M_{2,3},M_{0,2}] =  {1 \over z} \sinh(z M_{0,3}) \cosh(z M_{1,2})
\nonumber \\
&& [M_{1,3},M_{1,2}] = - M_{2,3}
\nonumber \\
&& [M_{1,3},M_{0,3}] = M_{0,1}
\nonumber \\
&& [M_{1,3},M_{0,1}] =   {1 \over z} \sinh(z M_{0,3}) \cosh(z M_{1,2})
\nonumber \\
&& [M_{1,2},M_{0,2}] = - M_{0,1}
\nonumber \\
&& [M_{1,2},M_{0,1}] = M_{0,2}
\nonumber \\
&& [M_{0,3},M_{0,2}] = M_{2,3}
\nonumber \\
&& [M_{0,3},M_{0,1}] = M_{1,3}
\nonumber \\
&& [M_{0,2},M_{0,1}] =  {1 \over z} \sinh(z M_{1,2}) \cosh(z M_{0,3})\label{1.1}
\end{eqnarray}
Observe that on the right hand sides we do not have linear functions
generators,  as in the Lie algebra case, but  some (analytic)
functions of them. However we still assume that the brackets are
antisymmetric and, as it is easy to show, that  Jacobi identity
holds.
 Note that in the limit $z\rightarrow0$ the algebra (\ref{1.1}) becomes the standard
 algebra $SO(3,1)$, and this is the reason for using the term $SO_q(3,1)$.

The  $SO(3,1)$ Lie algebra is the 2+1 dimensional de Sitter algebra
and it is  well known how to obtain the 2+1 dimensional Poincar\'e
algebra from it. First of all one has to single out the energy and
momentum generators of right physical dimension (note that the
generators $M_{\mu\nu}$ of (\ref{1.1}) are dimensionless): one
identifies three-momenta $P_\mu \equiv (E, P_i)$ ($\mu=1,2,3$,
$i=1,2$) as appropriately rescaled generators $M_{0,\mu}$ and then
one takes the  In\"om\"u--Wigner contraction limit. In the quantum
algebra case, the contraction is a bit more tricky, as one has to
convince oneself that after the contraction the structure one
obtains is still a quantum algebra. Such contractions has been first
discussed in Lukierski {\em et.\ al.}, 1991.

Let us try to  contract the algebra (\ref{1.1}). To this aim, since
momenta  are dimensionful, while the generators $M$ in (\ref{1.1})
are dimensionless, we must first rescale some of the generators by
an appropriate scale, provided by combination of dimensionful
constants present in definition of the parameter $z$
\begin{eqnarray}
&& E = \sqrt{\Lambda}\hbar\, M_{0,3}
\nonumber \\
&& P_i = \sqrt{\Lambda}\hbar\, M_{0,i}
\nonumber \\
&& M = M_{1,2}
\nonumber \\
&& N_i = M_{i,3}
\label{1.2}
\end{eqnarray}
Taking now into account the relation $z  \approx  \sqrt{\Lambda}
\hbar /  \kappa$, which holds for small $\Lambda$,   from
$$
[M_{2,3},M_{1,3}]  =  {1 \over z} \sinh(z M_{1,2}) \cosh(z M_{0,3})
$$
 we find
\begin{equation}\label{1.3}
[N_2,N_1]
= {\kappa \over \hbar \sqrt{\Lambda}} \sinh(\hbar \sqrt{\Lambda}/\kappa M)
\cosh( E/\kappa)
\end{equation}
Similarly from
$$
[M_{0,2},M_{0,1}] =  {1 \over z} \sinh(z M_{1,2}) \cosh(z M_{0,3})
$$
we get
\begin{equation}\label{1.4}
[P_2,P_1]
= {{\sqrt{\Lambda}\hbar \kappa}   }\, \sinh(\sqrt{\Lambda} \hbar / \kappa\, M)
\cosh( E/\kappa)
\end{equation}
Similar substitutions can be made in other commutators of
(\ref{1.1}). Now going to the contraction limit
$\Lambda\rightarrow0$, while keeping  $\kappa$ constant we obtain
the following algebra
\begin{eqnarray}
&& [N_i,N_j]  =  - M \epsilon_{ij}\, \cosh( E/\kappa)
\nonumber \\
&& [M,N_i] = \epsilon_{ij} N^j
\nonumber \\
&& [N_i,E] = P_i
\nonumber \\
&& [N_i,P_j] = \delta_{ij}\, {\kappa}\, \sinh( E/\kappa)
\nonumber \\
&& [M,P_i] = \epsilon_{ij} P^j
\nonumber \\
&& [E,P_i] = 0
\nonumber \\
&& [P_2,P_1] = 0
\label{1.5}
\end{eqnarray}
This algebra is called the three dimensional $\kappa$-Poincar\'e algebra (in the standard basis.)

Let us pause for a moment here to make couple of comments. First of
all, one easily sees that in the limit $\kappa\rightarrow\infty$
from the $\kappa$-Poincar\'e algebra algebra (\ref{1.5}) one gets
the standard Poincar\'e algebra. Second, we see that in this algebra
both the Lorentz and translation sectors are deformed. However,  in
the case of quantum algebras one is  free to change the basis of
generators in an arbitrary, analytic way (contrary to the case of
Lie algebras, where only linear trasformations of generators are
allowed.) It turns out that there exists such a change of the basis
that the Lorentz part of the algebra becomes classical (i.e.,
undeformed.) This basis is called bicrossproduct one, and the Doubly
Special Relativity model (both in 3 and 4 dimensions) based on such
an algebra is called DSR1.  In this basis the 2+1 dimensional
$\kappa$-Poincar\'e algebra looks as follows
\begin{eqnarray}
&& [N_i,N_j]  = -\epsilon_{ij}\, M
\nonumber \\
&& [M,N_i] = \epsilon_{ij} N^j
\nonumber \\
&& [N_i,E] = P_i
\nonumber  \\
&& [N_i,P_j] =   \delta_{ij}\, {\kappa\over 2}
 \left(
 1 -e^{-2{E/ \kappa}}
 + {\vec{P}\,{}^{ 2}\over \kappa^2}  \right) - \, {1\over \kappa}
P_{i}P_{j}
\nonumber \\
&& [M,P_i] = \epsilon_{ij} P^j
\nonumber \\
&& [E,P_i] = 0
\nonumber \\
&& [P_1,P_2] = 0 ~.
\label{1.6}
\end{eqnarray}

The algebra (\ref{1.6}) is nothing but the 2+1 dimensional analogue
of the algebra (\ref{1}) we started our discussion with. Thus we
conclude that in the case of 2+1 dimensional quantum gravity on de
Sitter space, in the flat space, i.e., vanishing cosmological
constant limit the standard Poincar\'e algebra is replaced by
(quantum) $\kappa$-Poincar\'e algebra.

It is noteworthy that in the remarkable paper by Freidel and Livine
$\kappa$-Poincar\'e algebra has been also found by direct
quantization of 2+1 gravity without cosmological constant, coupled
to point particles, in the weak gravitational constant limit. Even
though the structures obtained by them and the ones one gets from
contraction are very similar, their relation remains to be
understood.
\newline

Let me summarize. In 2+1 gravity (in the limit of vanishing
cosmological  constant) the scale $\kappa$, arises naturally. It can
be also shown that instead of the standard Poincar\'e symmetry we
have to do with the deformed algebra, with deformation scale
$\kappa$.

There is one interesting and important consequence of the emergence
of  $\kappa$-Poincar\'e algebra (\ref{1.6}). As in the standard case
this algebra can be interpreted both as the algebra of spacetime
symmetries and gauge algebra of gravity {\em and} the algebra of
charges associated with particle (energy momentum and spin.) It easy
to observe that this algebra can be interpreted as an algebra of
Lorentz symmetries of momenta if the momentum space is de Sitter
space of curvature $\kappa$. It can be shown that one can extend
this algebra to the full phase space algebra of a point particle, by
adding four (non-commutative) coordinates (see Kowalski-Glikman and
Nowak, 2003.) The resulting spacetime of the particle becomes the
so-called $\kappa$-Minkowski spacetime with the non-commutative
structure
\begin{equation}\label{7}
    [x_0, x_i] = - \frac1\kappa\, x_i
\end{equation}
On $\kappa$-Minkowski spacetime one can built field theory, which in
turn could be used to discuss phenomenological issues, mentioned in
the Introduction. In the next section I will show how, in a
framework of such a theory, one discovers the full power of quantum
$\kappa$-Poincar\'e algebra.

\section{Four dimensional field theory with curved momentum space}

As I said above $\kappa$-Poincar\'e algebra can be understood as  an
algebra of Lorentz symmetries of momenta, for the space of momenta
being the curved de Sitter space, of radius $\kappa$. Let us
therefore try to built the scalar field theory on such a space (see
also Daszkiewicz et.\ al.\ 2005.) Usually field theory is
constructed on spacetime, and then, by Fourier transform, is turned
to the momentum space picture. Nothing however prevents us from
constructing field theory directly on the momentum space, flat or
curved. Let us see how this can be done.

Let the space of momenta be de Sitter space of radius $\kappa$
\begin{equation}\label{8}
 -\eta_0^2 + \eta_1^2+ \eta_2^2+ \eta_3^2+ \eta_4^2 =\kappa^2,
\end{equation}
To find contact with $\kappa$-Poincar\'e algebra we introduce the coordinates on this space as follows
\begin{eqnarray}
{\eta_0} &=& -\kappa\, \sinh \frac{P_0}\kappa - \frac{\vec{P}\,{}^2}{2\kappa}\,
e^{  \frac{P_0}\kappa} \nonumber\\
\eta_i &=&   -P_i \, e^{  \frac{P_0}\kappa} \nonumber\\
{\eta_4} &=&  \kappa\, \cosh \frac{P_0}\kappa  - \frac{\vec{P}\,{}^2}{2\kappa}
\, e^{  \frac{P_0}\kappa},   \label{9}
\end{eqnarray}
Then one can easily check that the commutators of $P_\mu$ with
generators of Lorentz subgroup, $SO(3,1)$ of the full symmetry group
$SO(4,1)$ of (\ref{8}) form exactly the $\kappa$-Poincar\'e algebra
(\ref{1}).

In the standard, flat momentum space, case the action for free
massive scalar field has the form
\begin{equation}\label{10}
   S_0 = \int\, d^4P\, {\cal M}_0(P)\, \Phi(P)\, \Phi(-P)
\end{equation}
with ${\cal M}_0(P) = P^2- m^2$ being the mass shell condition. In
the case of  de Sitter space of momenta we should replace ${\cal
M}_0(P)$ with some generalized mass shell condition and also modify
somehow $\Phi(-P)$, because ``$-P$'' does not make sense on curved
space.

It is rather clear what should replace ${\cal M}_0(P)$. It should be
just the  Casimir of the algebra (\ref{1}). As a result of the
presence of the scale $\kappa$, contrary to the special relativistic
case, there is an ambiguity here. However since the Lorentz
generators can be identified with the generators of the $SO(4,1)$
algebra of symmetries of the quadratic form (\ref{8}), operating in
the $\eta_0$ -- $\eta_3$ sector, and leaving $\eta_4$ invariant it
is natural to choose the mass shell condition to be just (rescaled)
$\eta_4$, to wit
$$
m^2 = \kappa\, \eta_4 - \kappa^2
$$
so that
\begin{equation}\label{11}
  {\cal M}_\kappa(P) = \left( 2\kappa \sinh P_0/2\kappa\right)^2 - {\mathbf{P}^2}\, e^{P_0/\kappa} - m^2
\end{equation}
Eq.\ (\ref{11}) is the famous dispersion relation of DSR1. Notice
that it  implies that the momentum is bounded from above by
$\kappa$, while the energy is unbounded.

Let us now turn to the ``$-P$'' issue. To see what is to replace it
in  the theory with curved momentum space let us trace the origin of
it. In Special Relativity the space of momenta is flat, and equipped
with the standard group of motions. The space of momenta has the
distinguished point, corresponding to zero momentum. An element of
translation group $g(P)$ moves this point to a point of coordinates
$P$. This defines coordinates on the energy momentum space. Now we
{\em define} the point with coordinates $S( P)$ to be the one
obtained from the origin by the action of the element $g^{-1}(P)$.
Since the group of translations on flat space is an abelian group
with addition, $S(P) = -P$.

Now, since in the case of interest the space of momenta is de Sitter
space,  which is a maximally symmetric space, we can repeat exactly
the same procedure. The result, however is not trivial now, to wit
\begin{equation}\label{12}
    S(P_0) = - P_0, \quad S(P_i )= - e^{P_0/\kappa}\, P_i
\end{equation}
Actually one can check that the $S$ operator is in this case nothing
but  the antipode of $\kappa$-Poincar\'e quantum algebra. Thus we
can write down the action for the scalar field on curved momentum
space as
\begin{equation}\label{13}
   S_\kappa = \int\, d^4P\, {\cal M}_\kappa(P)\, \Phi(P)\, \Phi(S(P))
\end{equation}

De Sitter space of momenta has the ten-dimensional group of
symmetries,  which can be decomposed to six ``rotations'' and four
remaining symmetries, forming the deformed $\kappa$-Poincar\'e
symmetry (\ref{1}). We expect therefore that the action (\ref{13})
should, if properly constructed, be invariant under action of this
group. We will find that this is indeed a case, however the story
will take an unexpected turn here: the action will turned out to be
invariant under action of the {\em quantum group}.

Let us consider the four-parameter subgroup of symmetries, that in
the  standard case would correspond to spacetime translation. It is
easy to see that in the standard case the  translation in spacetime
  fields is in the one-to-one correspondence with the phase transformations of the momentum
  space ones. This suggests that the ten parameter group of Poincar\'e symmetries in
  space-time translates into six parameter Lorentz group plus four independent phase
  transformations in the momentum space, being representations of the same algebra.

Using this insight let us  turn to the case at hands. Consider first
the   infinitesimal phase transformation in energy
direction\footnote{Note that since the function ${\cal M}$ is real,
$\delta_0{\cal M}_\kappa = \delta_i{\cal M}_\kappa =0$.} (to
simplify the notation I put $\kappa=1$)
\begin{equation}\label{V.4}
    \delta_0 \Phi(P_0, \mathbf{P}) = i \epsilon\, P_0 \Phi(P_0, \mathbf{P}),
\end{equation}
where $\epsilon$ is an infinitesimal parameter. It follows that
\begin{equation}
\delta_0 \Phi(S(P_0), S(\mathbf{P})) = i \epsilon\, S(P_0)\,
\Phi(S(P_0), S(\mathbf{P})) = -i \epsilon\, P_0\,
\Phi(S(P_0), S(\mathbf{P}))
\end{equation}
and using Leibniz rule we easily see that the action is indeed
invariant. Let us now consider the phase  transformation in the
momentum direction. Assume that in this case
\begin{equation}\label{V.5}
    \delta_i \Phi(P_0, \mathbf{P}) = i \epsilon\, P_i\, \Phi(P_0, \mathbf{P}).
\end{equation}
But then
$$
\delta_i \Phi(S(P_0), S(\mathbf{P})) = i \epsilon\, S(P_i)\, \Phi(S(P_0), S(\mathbf{P})) =$$
\begin{equation} -i \epsilon \, e^{P_0}\,  P_i\, \Phi(S(P_0), S(\mathbf{P}))
\end{equation}
and the action is not invariant, if we apply Leibniz rule.

The way out of this problem is to replace the Leibniz rule by the
co-product one.  To this end we take
$$
\delta_i\left\{ \Phi(P_0, \mathbf{P}) \Phi(S(P_0),
S(\mathbf{P})) \right\}
    \equiv $$ $$\delta_i\left\{ \Phi(P_0, \mathbf{P})\right\} \Phi(S(P_0),
    S(\mathbf{P}))
    + \left\{e^{-P_0} \Phi(P_0, \mathbf{P})\right\}\,
    \delta_i\left\{\Phi(S(P_0), S(\mathbf{P})) \right\} =0
$$
i.e., we generalize Leibniz rule by multiplying $\Phi(P_0, \mathbf{P})$
in the second term by $e^{-P_0}$. Note that such definition is consistent with
the fact that the fields are commuting, because
$$
  \delta_i\left( \Phi(S(P_0), S(\mathbf{P})) \Phi(P_0, \mathbf{P}) \right) =
  $$
   $$ \left( i \epsilon\, S( P_i) + i \epsilon\, e^{-S(P_0)}\, P_i\right)\Phi(S(P_0),
   S(\mathbf{P})) \Phi(P_0, \mathbf{P}) =0.
$$
We see therefore that in order to make the action invariant with
respect to infinitesimal  phase transformations one must generalize
the standard Leibniz rule to the non-symmetric co-product one.

The rule of how an algebra acts on (tensor) product of objects is
called the co-product,  and denoted by $\Delta$. If Leibniz rule
holds the coproduct is trivial $\Delta\delta = \delta\otimes1 +
1\otimes\delta$. Quantum groups can be characterized by the fact
that Leibniz rule is generalized to a non-trivial coproduct rule. We
discovered that in the case of $\kappa$-Poincar\'e algebra it takes
the form
\begin{equation}\label{14}
   \Delta \delta_0 = \delta_0\otimes1 + 1\otimes\delta_0, \quad \Delta \delta_i
   = \delta_i\otimes1 + e^{-P_0}\otimes\delta_i
\end{equation}
One can check that, similarly, the co-product for rotational part
of the symmetry algebra is also non-trivial. The presence of
non-trivial co-product in the algebraic structure of DSR theory has,
presumably, far reaching consequences for particle kinematics. I
will return to this point below.

\section{DSR phenomenology}

DSR has  emerged initially from the quantum gravity phenomenology
investigations, as a phenomenological theory, capable of describing
possible future observations disagreeing with predictions of Special
Relativity. Two of these effects, the possible energy dependence of
the speed of light, which could be observed by GLAST satellite, and
the mentioned already, possible violation of the GZK cutoff, which
could be confirmed by Pierre Auger Observatory were quite
extensively discussed in the literature. Let me now briefly describe
what would be the status of this (possible) effects vis a vis the
approach of DSR I have been analyzed above\footnote{It should be
stressed that DSR has been originally proposed as an idea, not a
formally formulated theory, and therefore it may well happen that
the particular realization of this idea described above could be
replaced by another one in the future.}.

The prediction of energy dependence of the speed of light is based
on the rather naive observation that since in (some formulations of)
DSR the dispersion relation is being deformed, the formula for
velocity $v = \partial E/\partial p$ gives, as a rule, the result
which differs from this of Special Relativity. It turns out however
that this  conclusion may not stand if the effects of
non-commutative spacetime are taken into account.

In the classical theory the non-commutativity is replaced by the
nontrivial structure of the phase space of the particle, and, as in
the standard case, one calculates the three velocity of the particle
as the ratio of $\dot x = \{ x, H\}$ and $\dot t = \{ t, H\}$: $v =
\dot x/\dot t$. Then it can be generally proved that the effect of
this nontrivial phase space structure cancels neatly the effect of
the modified dispersion relation (see Daszkiewicz et.\ al.\ 2004 for
details.) Thus, in the framework of this formulation of DSR, the
speed of massless particles is always 1, though there are deviations
from the standard Special Relativistic formulas in the case of
massive particles. However the leading order corrections are here of
order of $m/\kappa$, presumably beyond the reach of any feasible
experiment.
\newline

Similarly one can argue that deviations from the GZK cutoff should
be negligibly small in any natural DSR theory. The reasoning goes as
follows (similar argument can be found in Amelino-Camelia 2003.)
Consider {\em experimental} measurement of the threshold energy for
reaction $p+\gamma = p + \pi^0$, which is one of the relevant ones
in the ultra high energy cosmic rays case, but details are not
relevant here. To measure this energy we take the proton initially
at rest and bombard it by more and more energetic photons. At some
point, when the photon energy is of order of $E^0_{th}=145$ MeV, the
pion is being produced. Note that the threshold energy is just
$E^0_{th}$, exactly as predicted by Special Relativity, and the
corrections of DSR (if any) are much smaller than the experimental
error bars $\Delta E^0_{th}$. Thus whichever kinematics is the real
one we have the robust result for the value of the threshold energy.

Now there comes the major point. Since DSR respects the Relativity
Principle by definition, we are allowed to boost the photon energy
down to the CMB energy (this cannot be done in the Lorentz breaking
schemes, where the velocity of the observer with respect to the
ether matters), and to calculate the value of the corresponding
rapidity parameter. Now we boost the proton with the same value of
rapidity, using the DSR transformation rules, and check what is the
modified threshold. Unfortunately, the leading order correction to
the standard Special Relativistic transformation rule would be of
the form $\sim \alpha E_{proton}/\kappa$, where $E_{proton}$ is the
energy of the proton after boost, and $\alpha$ is the numerical
parameter, fixed in any particular formulation of DSR. It is natural
to expect that $\alpha$ should be of order 1, so that in order to
have sizeable effect we need $\kappa$ of order of $10^{19}$ eV,
quite far from the expected Planck scale\footnote{Note that in this
reasoning we do not have to refer to any particular DSR kinematics,
the form of energy-momentum conservation etc. The only input here is
the Relativity Principle.}. One may contemplate the idea that since
the proton is presumably, from the perspective of the Planck scale
physics, a very complex composite system, we do not have to do here
with ``fundamental'' $\kappa$, but with some effective one instead,
but then this particular value should be explained (it is curious to
note in this context that, as observed in Amelino-Camelia 2003,
$10^{19}$ eV is of order of the geometric mean of the Planck energy
and the proton rest mass.) However the conclusion for now seems
inevitably be that with the present formulation of DSR, the
explanation of possible violation of GZK cutoff offered by this
theory is, at least, rather unnatural.

\section{DSR -- facts and prospects}

Let me summarize. Above I stressed two facts that seem to be essential features of DSR theory.

First (quantum) gravity in 2+1 dimensions coupled to point particles
is  just a DSR theory. Since the former is rather well understood,
it is a perfect playground for trying to understand better the
physics of the latter. In 3+1 dimensions the situation is much less
clear. Presumably, DSR emerges in an appropriate limit of (quantum)
gravity, coupled to point particles, when the dynamical degrees of
freedom of gravitational field are switched off, and only the
topological ones remain. However, it is not known, what exactly this
limit would be, and how to perform the limiting procedure in the
full dynamical theory. There is an important insight, coming from
algebraic consideration, though. In 3+1 dimension one can do almost
exactly the same procedure as the one, I presented for the 2+1 case
above. It suffices to replace the $SO_q(3,1)$ group with the
$SO_q(4,1)$. It happens however that in the course of the limiting
procedure one has to further rescale the generators corresponding to
energy and momentum. The possible rescalings are parametrized  by
the real, positive parameter $r$: for $r>1$ the contraction does not
exist, for $0<r<1$ as the result of contraction one gets the
standard Poincar\'e algebra, and only for one particular value $r=1$
one finds $\kappa$-Poincar\'e algebra. This result is not understood
yet, and, if DSR is indeed a limit of gravity, gravity must tell us
why one has to choose this particular contraction.

Second, as I explained above there is a direct interplay between
non-trivial co-product and the fact that momentum space is curved.
In addition, curved momentum space naturally implies non-commutative
space-time. While the relation between these three properties of DSR
theory has been well established, it still requires further
investigations.

The presence of the non-trivial co-product in DSR theory has its
direct consequences for particle kinematics. Namely the co-product
can be understood as a rule of momentum composition. This fact has
been again well established in the 2+1 dimensional case. However the
3+1 situation requires still further investigations. The main
problem is that the co-product composition rule is not symmetric:
the total momentum of the system (particle$_1$ + particle$_2$) is
not equal, in general,  to the of the total momentum of
(particle$_2$ + particle$_1$) one. This can be easily understood in
2+1 dimension if one thinks of particles in terms of their
worldlines, and where the theory takes care of the worldlines'
braiding. In 3+1 dimensions the situation is far from being clear,
though. Perhaps a solution could be replacing holonomies that
characterize particles in 2+1 one dimensions by surfaces surrounding
particles in 3+1 dimensions. If this is true, presumably the theory
of gerbes will play a role in DSR (and gravity coupled with
particles, for that matter.)

Related to this is the problem of ``spectators''. If the co-product
rule is indeed correct,  any particle would feel non-local influence
of other particles of the universe. This means in particular, that
LSZ theorem of quantum field theory, which requires the existence of
free asymptotic states, presumably does not hold in DSR, and thus
the whole of basic properties of QFT will have to be reconsidered.

Arguably one of the most urgent problems of DSR is the question
``what is the momentum?'' Indeed, as I mentioned above, in the
$\kappa$-Poincar\'e case one has a freedom do redefine momentum and
energy by any function of them and the $\kappa$ scale, restricted
only by the condition that in the limit $\kappa\rightarrow\infty$
they all reduce to the standard momenta of special relativity. In
particular some of them might be bounded from above, and some not.
For example in DSR1 momentum is bounded from above and energy is
not, in another model, called DSR2 both energy and momentum are
bounded, and there are models in which neither is. Thus the question
arises as to which one of them is physical? Which momentum and
energy we measure in our detectors?

There is a natural answer to this question. Namely, the physical
momentum is the charge that couples to gravity. Indeed if  DSR is an
emergent theory, being the limit of gravity, the starting point
should be, presumably, gravity coupled to particles' Poincar\'e
charges in the canonical way.

To conclude. There seem to be an important and deep interrelations
between developments in quantum gravity and understanding of DSR.
Proper control over semiclassical quantum gravity would provide an
insight into the physical meaning and relevance of DSR. And vice
versa, DSR, being a possible description of ultra high energetic
particle behavior will perhaps become a workable model of quantum
gravity phenomenology, to be confronted with future  experiments.

\section*{Acknowledgements}

This work  is partially supported by the KBN grant 1 P03B 01828.

\end{document}